\documentclass[useAMS,usenatbib,usegraphicx]{mn2e}

\usepackage{lscape}

\newcommand{\kms}{\hbox{km s$^{-1}$}}

\newcommand{\vsini}{\hbox{$v$\,sin\,$i$}}

\newcommand{\degs}{$\degr$}
\newcommand{\chisq}{$\chi^{2}$}

\newcommand{\hd}{\hbox{HD 189733}}
\newcommand{\hdb}{\hbox{HD 189733b}}

\title[Search for molecules in the atmosphere of \hdb]{A search for molecules in the atmosphere of \hdb}

\author[J.R.~Barnes et al.]
{J.R.~Barnes$^1$\thanks{E-mail: j.r.barnes@herts.ac.uk} 
 Travis~S.~Barman$^2$,
 H.R.A.~Jones$^1$,
 R.J.~Barber$^3$,
 Brad~M.S.~Hansen$^4$,
\newauthor 
 L.~Prato$^2$,
 E.L. Rice$^4$,
 C.J.~Leigh$^5$, 
 A.~Collier~Cameron$^6$,
 D.J.~Pinfield$^1$ \\
$^1$ Centre for Astrophysics Research, University of Hertfordshire, Hertfordshire AL10 9AB. UK \\ 
$^2$ Lowell Observatory, Planetary Research Centre, 1400 West Mars Hill Road, Flagstaff, AZ 86001. USA \\
$^3$ Departments of Physics and Astronomy, University College London, London WC1E 6BT. UK. \\
$^4$ Department of Physics and Astronomy, and Institute of Geophysics and Planetary Physics, \\~~~University of California Los Angeles, Los Angeles, CA 90095, USA \\
$^5$ Astrophysics Research Institute, Liverpool John Moores University, Birkenhead CH41 1LD. UK \\
$^6$ SUPA, School of Physics and Astronomy, University of St Andrews, Fife KY16 9SS. UK \\
}

\begin{document}

\date{Accepted xxxx. Received 2009}


\maketitle

\label{firstpage}

\begin{abstract}

{We use signal enhancement techniques and a matched filter analysis to search for the K band spectroscopic absorption signature of the close orbiting extrasolar giant planet, \hdb. With timeseries observations taken with NIRSPEC at the Keck II telescope, we investigate the relative abundances of H$_2$O and carbon bearing molecules, which have now been identified in the dayside spectrum of HD 189733b.  We detect a candidate planet signature with a low level of significance, close to the $\sim 153$ kms$^{-1}$ velocity amplitude of \hdb. However, some systematic variations, mainly due to imperfect telluric line removal, remain in the residual spectral timeseries in which we search for the planetary signal. Using principal components analysis, the effects of this pattern noise may be reduced. Since a balance between the optimum systematic noise removal and minimum planetary signal attenuation must be struck, we find that residuals, which are able to give rise to candidate planet signatures, remain. The robustness of our candidate signature is therefore assessed, enabling us to conclude that it is not possible to confirm the presence of any planetary signal which appears at $F_p/F_*$ contrasts deeper than the 95.4 per cent confidence level. Our search does not enable us to detect the planet at a contrast ratio of \hbox{$F_p/F_*$ = 1/1920} with 99.9 per cent confidence.

Finally, we investigate the effect of model uncertainties on our ability to reliably recover a planetary signal. The use of incorrect temperature, model opacity wavelengths and model temperature-pressure profiles have important consequences for the least squares deconvolution procedure that we use to boost the S/N ratio in our spectral timeseries observations. We find that mismatches between the empirical and model planetary spectrum may weaken the significance of a detection by $\sim$~30\,-\,60 per cent, thereby potentially impairing our ability to recover a planetary signal with high confidence.}

\end{abstract}


\section{Introduction}

{The field of exoplanet spectroscopy has advanced rapidly since the first observations of secondary eclipse events enabled the brightness temperatures of transiting planets to be determined \citep{charbonneau05tres1,deming05hd209458}, thereby confirming heating due to stellar irradiation}. Thanks largely to the Spitzer Space Telescope \citep{houck04irs,fazio04irac,rieke04mips}, the spectral energy distribution of a number systems has now been estimated (e.g. see \cite{burrows08cegp}). It has recently been suggested that close orbiting extrasolar giant planets (CEGPs) can be divided into two possible sub-groups, one which exhibits absorption spectra and one which exhibits emission features due to the presence of a stratosphere \citep{burrows08cegp,fortney08unified}. Individual planetary atmospheres are no doubt more complex with \cite{burrows08cegp}, for example, preferring to use parameterisations of the degree of stratospheric absorption and heat redistribution. Nevertheless, HD 209458b appears to fall broadly into the latter category \citep{knutson08hd209458b} while the spectral energy distribution of \hdb\ is found to be well fit by models where no stratosphere forms \citep{barman08hd189733b}.

Although planetary signatures are easier to {\em detect} at mid-infrared wavelengths, {\em characterisation} of atmospheres is easier at shorter wavelengths where observational sensitivities are sufficient to probe the large amplitude signatures of species such as H$_2$O and carbon bearing molecules. The first evidence for water and organic molecules in the atmospheres of CEGPs has come from space based broadband photometric observations and low resolution transit spectroscopy. While \cite{barman07} and \cite{tinetti07nature} claimed independent detections of H$_{2}$O at different wavelengths, \cite{swain08methane} have now identified CO and CH$_4$ in transit spectra which probe the terminator of the planet HD 189733b. \cite{grillmair08dayside} reported a detection of water in the dayside spectrum through Spitzer space telescope observations of secondary eclipse events below \hbox{7.5 \micron}. Most recently NICMOS/HST data have been used to infer the additional presence of CO and CO$_2$ in the dayside spectra of HD 189733b (\citealt{swain09dayside}, hereafter S09).

Rather than attempting to identify molecules from their broadband spectral signatures, we present a method which attempts to identify {spectral structure} through a statistical examination of the many thousand individual transitions found in a typical high resolution (R $\sim$ 25,000\,-\,50,000) planetary spectrum. Specifically, in order to detect the faint planetary signal, spectral deconvolution is applied to individual spectra in a timeseries in order to derive mean absorption profiles with boosted S/N ratios. Modelling the phase dependent contrast ratio and radial velocity motion enables the maximum planet/star contrast ratio and velocity amplitude to be measured (from which the true mass of the planet may also be determined). Crucially, since the planet need not be transiting, this spectroscopic method increases the sample of planets which can potentially be studied with current instrumentation. With the aim of extending space based mid-infrared measurements of planet/star contrast ratios into the near infrared, we search for the CEGP signature, manifested as H$_2$O, CO and CO$_2$ absorption in high resolution K band spectroscopic timeseries observations of \hbox{HD 189733} (K1V-K2V).

\begin{table*}
\caption{Keck/NIRSPEC observations of \hd\ for UT 2008 June 15 and 22. }
\protect\label{tab:journal}
\vspace{5mm}
\begin{center}
\begin{tabular}{lccccc}
\hline
UT Date		& UT start of	& UT start of	& Time per	  & Number of		& Number of 	  \\
		& first frame	& last frame	& exposure [secs] & co-adds per frame	& observations	  \\
\hline
2008 June 15 	& 08:36:29	& 15:18:33	&	5	  & 	12		& 219		\\
2008 June 22 	& 10:43:37	& 15:08:46	&	5	  & 	12		& 154		 \\
\hline
\end{tabular}
\end{center}
\end{table*}

\section{Data reduction}

\subsection{Observations}
K band observations of \hd\ were secured with NIRSPEC \citep{mclean98nirspec} at the Keck II Telescope on UT 2008 June 15 and June 22. Respectively, a total of 219 and 154 spectra were recorded using a $1024\,\times\,1024$ InSb Aladdin-3 array. With the NIRSPEC-7 blocking filter, a wavelength span of 2.0311 \micron\,-\,2.3809 \micron\ was achieved with a slit width of 0.432\arcsec, giving a resolution of \hbox{R $\sim$ 25, 000}. Our 60 sec exposures comprised of 12 coadds, each of 5 secs duration. The observations are summarised in Table \ref{tab:journal}. The seeing was mostly good at around 0.6\,-\,0.7\arcsec\ although observations were plagued by cloud for a period on June 15.

\subsection{Data extraction}
Pixel to pixel variations were corrected for each frame using flat-field exposures taken with an internal tungsten reference lamp. The worst cosmic ray events were removed at the pre-extraction stage using the Starlink {\sc figaro} routine {\sc bclean} \citep{shortridge93figaro}. Since we chose not to use an ABBA nodding sequence, in order to maximise stability in the K band, we carried out the same extraction procedure as detailed in \citet{barnes07b} for previous observations of \hd. The spectra were extracted using {\sc echomop}'s implementation of the optimal extraction algorithm developed by \citet{horne86extopt}. {\sc echomop} rejects all but the strongest sky lines \cite{barnes07b} and propagates error information based on photon statistics and readout noise throughout the extraction process. 

\begin{figure}
\begin{center}
   \includegraphics[width=75mm,angle=270]{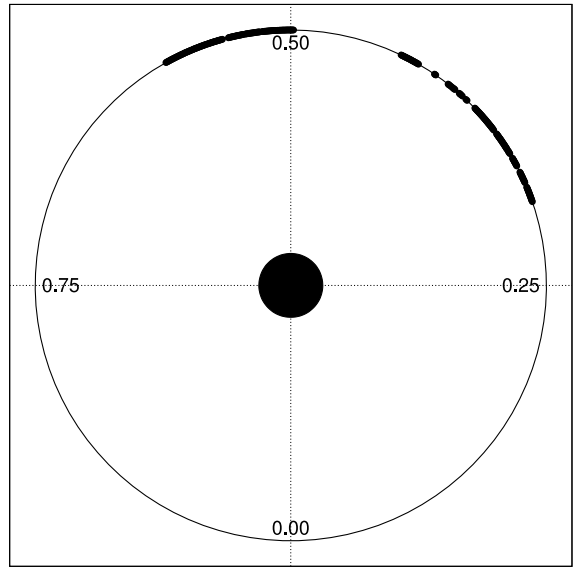} 
   \caption{HD 189733b orbital phase diagram. {Observations made on June 15 cover phases $\phi = 0.303\,-0.429$ while observations made on June 22 cover $\phi = 0.498\,-\,0.581$. Observations in the range $\phi = 0.498\,-\,0.517$ were not used in the subsequent analysis since the planet is eclipsed during these phases.} Only the phases used for analysis are shown.}
\end{center}
\protect\label{fig:phases}
\end{figure}

In \cite{barnes07b}, we reported observations of \hd\ which were made in conditions of variable cloud and seeing. This resulted in a spectral timeseries in which the S/N ratio of individual spectra varied greatly. We found that rejecting the lowest S/N frames increased the sensitivity of our method. We thus carried out a similar rejection procedure, set at an arbitrary S/N = 350 to reject the lowest S/N ratio frames which formed a distinct distribution separate from those frames which were observed in the best seeing conditions. A total of 63 frames were rejected from observations made on June 15 (S/N ratios of the rejected frames were $245 \pm 127$) while none were rejected from observations made on June 22; a clear reflection of the more variable conditions prevailing on the first night. The S/N ratios of the residual timeseries were $497 \pm 125$ and $570 \pm 77$ on June 15 and 22 respectively. The phases of observations used during the subsequent analysis stages are represented visually in Fig. 1.

\subsection{Residual spectra}

Following the procedures outlined in \cite{barnes07b} and \cite{barnes08}, we attempt to extract the planetary signature from timeseries spectra by removing the dominant spectral contributions; namely the stellar spectrum and the telluric lines. A master spectrum is generated by combining all observed frames on a given night after nearest-pixel alignment of each individual order in each spectrum to minimise blurring. {The master template is subtracted from each spectrum in turn after shifting, scaling and blurring/sharpening (again on an order to order basis). This latter procedure is achieved by calculating the first to fourth order derivatives of the template spectrum. The master template is then scaled to each observed spectrum by using a spline with a fixed number of knots. Scale factors for the derivatives are similarly calculated. A model of each observed spectrum can thus be calculated by means of a Taylor expansion of the template spectrum. The procedure is described in detail in Appendix A of \cite{cameron02upsand}}. Since we observe \hdb\ at phases close to phase $\phi = 0.5$ when the planet spectrum is Doppler shifted on the steepest part of the radial velocity curve, the master template spectrum contains only a very weak, blurred out copy of the planetary spectrum. Hence subtraction of the template does not significantly attenuate the planetary signature.

\subsection{Residual pattern noise removal}
{The resulting residual timeseries should thus ideally only contain noise and a copy of the planetary spectrum. However, residual pattern noise remains in the subtracted timeseries. Examination indicates that the dominant patterns are time variant residuals which are introduced when subtracting the scaled master template during the procedure outlined in \S 2.3. Although this procedure is effective at removing the stellar and telluric lines, variations between the strength of these lines throughout the night (due to changing airmass and hence telluric strength) are sufficient to preclude consistent results over timescales of the variations. 

In addition to pattern noise induced by telluric effects, noise may arise from time variable fringing effects, as found by \cite{brown02hd209458b} for K band NIRSPEC observations for example. To remove these two effects, \cite{brown02hd209458b} carried out a two stage procedure in the form of a regression and singular value decomposition filtering procedure on their spectra. Our implementation of the Taylor expansion algorithm, mentioned above and in detail in \cite{cameron02upsand}, is analogous to the regression step. In a similar procedure to the singular value decomposition implemented by \cite{brown02hd209458b}, the remaining pattern nose can be removed by calculating the correlation matrix of the time variations in individual spectral bins for our set of residual spectra. The eigenvectors of the correlation matrix, which account for the largest fraction of the variance, are the principal components of the residual spectra. By subtracting only the principal components, any remaining pattern noise can be further removed (see Appendix B of \citealt{cameron02upsand} for further details). A balance must be struck between removing pattern noise and maintaining information in the residual spectra. We found that shifts of 1-2 pixels during a typical night of observations could be attributed to points during the night where the instrumental configuration was slightly modified {to enable observations of a star for another project}. We thus applied principal components analysis to continuous blocks (i.e. between observations of the other star, where the 1-2 pixel shifts occurred) of HD 189733 observations and found that 2 to 3 components were necessary to remove the remaining pattern noise in the data without attenuating the planetary signal (see \S 4).

}

\section{Deconvolution and planetary models }

In order to extract the planetary signature from the corrected residual spectra, we used least squares deconvolution \citep{donati97zdi} which requires the use of a model spectrum to describe the strengths (normalised profile depths) and wavelength positions of the strongest planetary opacities over the wavelength range of our observations. Our implementation of the algorithm \citep{barnes98aper} propagates errors from the input spectra and has been used in reflected light searches in the optical by \citet{cameron99tauboo,cameron02upsand} and \citet{leigh03a,leigh03b}. For each residual spectrum a deconvolved profile is obtained, potentially containing a copy of the Doppler shifted planetary profile which can then be detected owing to the effective boost in S/N ratio. Typical boosts in S/N ratio of a few times to a few tens of times are achieved since several hundred to several thousand planetary absorption lines are used to deconvolve the planetary signature. The 2.04\,-\,2.06 \micron\ and 2.36\,-\,2.38 wavelength ranges are omitted in all the following analyses due to the strong telluric features which dominate these regions of the spectra.

We have generated several models to represent the emergent spectrum of HD 189733b. {Our standard \hdb\ model, an updated version of the model detailed in \S 3.2 of \cite{barnes07b}, is generated for an atmosphere with solar metallicity, a temperature, \hbox{T = 1250 K} and surface gravity, \hbox{log $g = 1.33$ ms$^{-1}$}.} For a {detailed} description of the model opacities and setup see \citet{ferg05}, \citet{barman01} and \citet{bha05} BHA05. {The most recent models and their ability to fit \hdb\ observations made with the Spitzer Space Telescope are discussed at length in \cite{barman08hd189733b}.} A number of further models with adjusted temperature pressure profiles and relative chemical abundances will be discussed in the following sections. The S/N ratio after deconvolution with the standard model yields timeseries with mean profile S/N ratios of $7520$ and $9450$ on June 15 and 22 respectively. The mean deconvolved wavelength of the spectra depends on the wavelength dependent count rate and the relative strengths of the absorption lines. For the standard model, \hbox{$\lambda_{mean} = 2.19$ \micron}. As reported in \S 4, the mean wavelength may change slightly when the line list used for deconvolution is modified.


\begin{figure*}
 \begin{center}
   \begin{tabular}{cc}

      \vspace{10mm} \\
      \includegraphics[width=70mm,bbllx=113,bblly=114,bburx=397,bbury=397,angle=0]{barnes_hd189733_2009_fig2.ps} \hspace {5mm} &
      \hspace {5mm}
      \includegraphics[width=70mm,bbllx=113,bblly=114,bburx=397,bbury=397,angle=0]{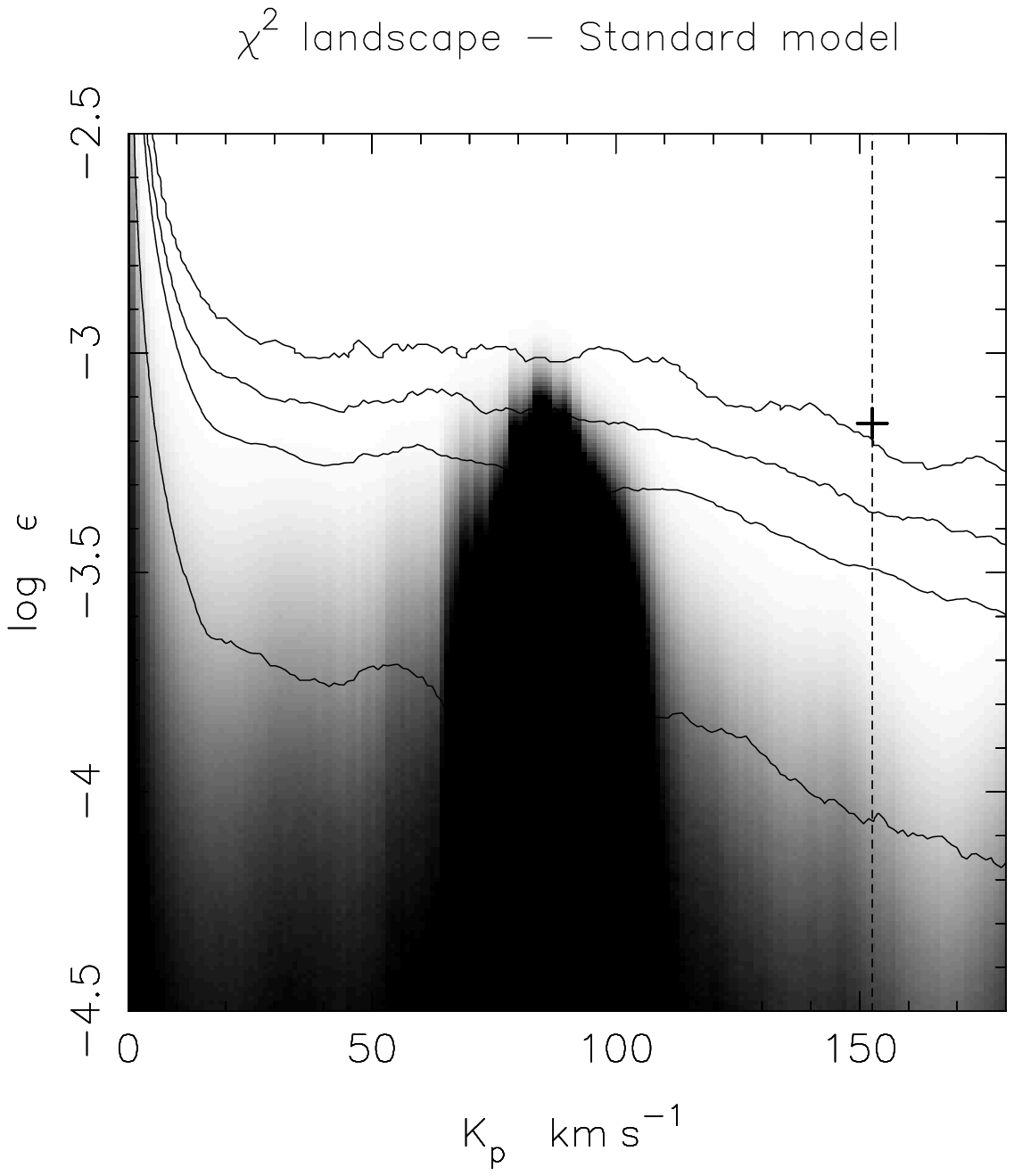} \\
      \vspace{10mm} \\

      \vspace{10mm} \\
      \includegraphics[width=70mm,bbllx=113,bblly=114,bburx=397,bbury=397,angle=0]{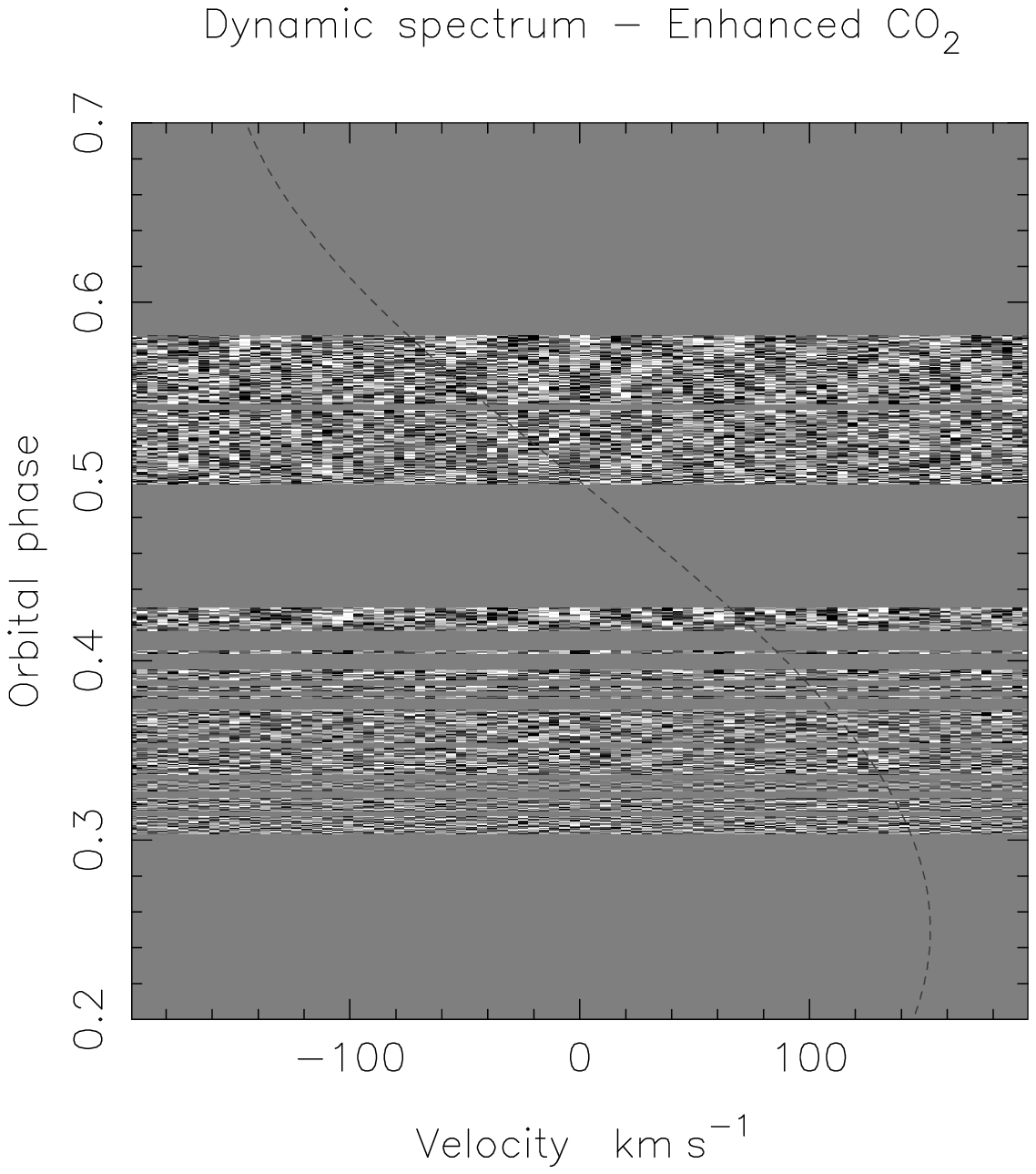} \hspace {5mm} &
      \hspace {5mm}
      \includegraphics[width=70mm,bbllx=113,bblly=114,bburx=397,bbury=397,angle=0]{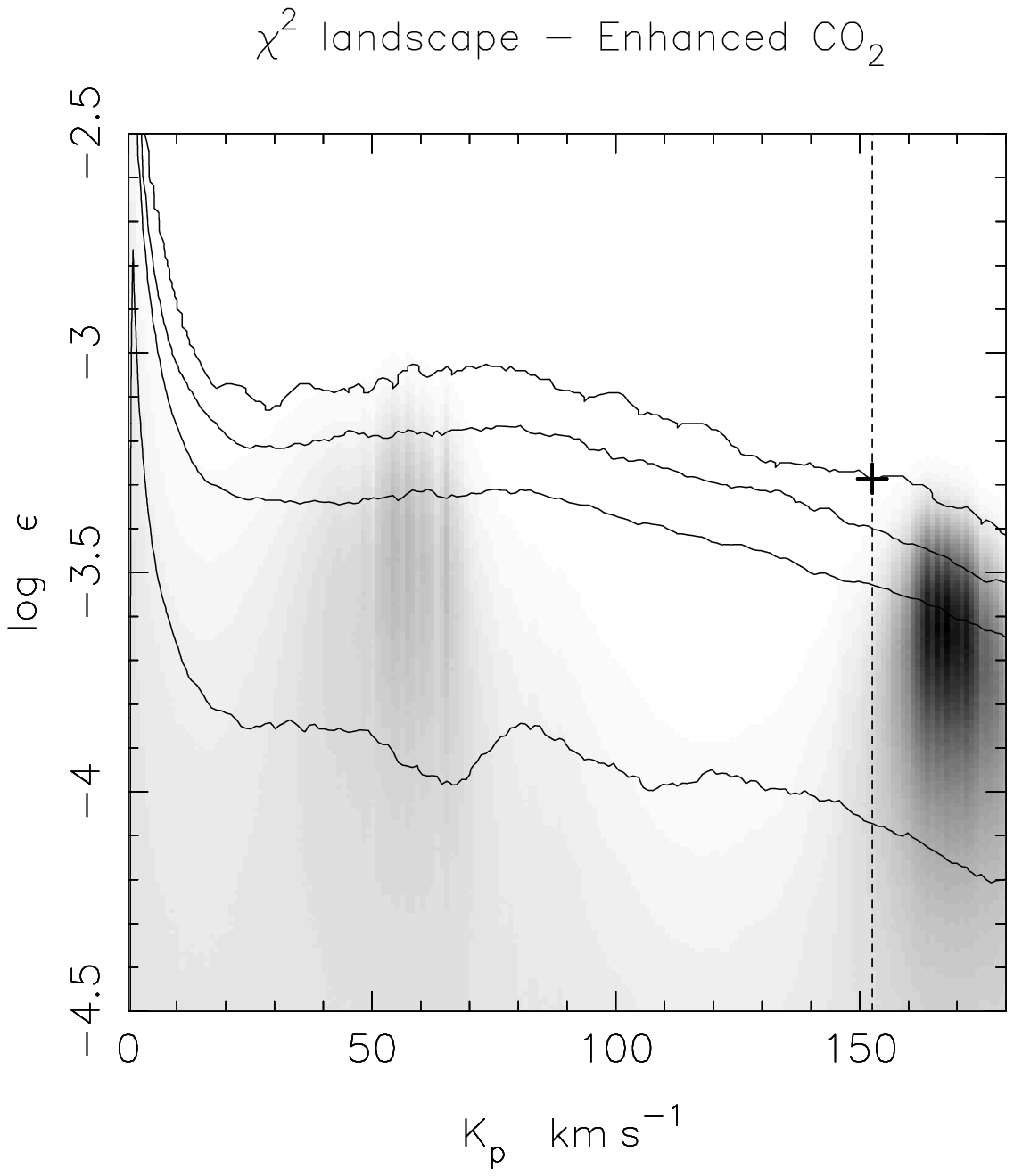} \\
      \vspace{10mm} \\

   \end{tabular}
 \end{center}
\caption{Phased deconvolved timeseries spectra of HD 189733b and corresponding 2-dimensional \chisq\ landscape plots for analysis using the standard model (top) and enhanced CO$_2$ model (bottom). Left: The dashed line in the timeseries plots represents the position of the planet as a function of phase. The black and white levels are set at the 1.5$\sigma$ level of the timeseries. Right: $\chi^2$ plot for matched filter combinations of maximum contrast ratio, log$_{10}$($\epsilon_0$) vs $K_p$. {Observations covering phases $\phi = 0.498\,-\,0.517$ are shown for completeness but were not used in the analysis since the planet undergoes eclipse at these phases.} Black and white represent the best and worst improvements in \chisq\ respectively. The recovered enhancements in $\chi^2$ are {\em likely} caused mainly by systematics in the residual timeseries spectra. No enhancement in $\chi^2$ is measured at the known velocity amplitude, $K_p = 152.6$ \kms, marked by the dashed line. Upper confidence levels of 63.8, 95.4, 99 \& 99.9 per cent (top to bottom solid lines) are plotted. The bold + symbols mark the mean value of the contrast ratio over the range of observations for the standard model (top right) and the corresponding mean value of the contrast ratio from the NICMOS/HST S09 observations (bottom) and are \hbox{log$_{10}(\epsilon_0)$ = -3.163 \& -3.286} ($F_p/F_* = 1/1460~\&~1/1930$) respectively.}
\protect\label{fig:planet_signal_kband}
\end{figure*}

\section{Results}

A matched Gaussian filter approach is used to search for the planetary signal, where a model describing the radial velocity shift and planet/star contrast phase function is used to search for the deconvolved absorption profile of the planet in the spectral timeseries \citep{cameron02upsand,barnes07a,barnes07b,barnes08}. Pairs of maximum contrast ratio ($\epsilon_0$) vs velocity amplitude ($K_p$) are used in a two dimensional \chisq\ search to find the combination which yields the best improvement in \chisq. The significance of candidate enhancements in \chisq\ are assessed by a bootstrap procedure which randomises the order of the data within each night of observations \citep{cameron02upsand}. This process, carried out several thousand times, scrambles any planetary signature, but enables the data to retain the ability to give false \chisq\ enhancements due to systematics which may remain in the data above the photon noise. Reliable confidence levels can thus be plotted on the 2-parameter \chisq\ landscapes of \hbox{log$_{10}$($\epsilon_0$)} vs ($K_p$) \chisq.

Calibration of contrast ratios is achieved by injecting a ``fake'' model planet spectrum into the timeseries and then recovering the signal using our matched filter method. The fake planetary spectrum is injected with known $\epsilon_0$ and $K_p$ after extraction of the spectra and before any of the subsequent steps described above are carried out. In this way, we are also able to assess our ability to correctly recover a planetary signature \citep{barnes07b,barnes08}. We find that for a planet recovered with high significance, there may be a slight shift in the recovered $K_p$ velocity. This is most likely due to some removal of planetary signature when subtracting the template star and during principal components analysis. During these procedures, we chose parameters which strike a balance between removing residuals while not significantly affecting the planet signal. Essentially, an absorption signature located at phases close to $\phi = 0.25$ will be attenuated most, since, in this region the planet shows the smallest radial velocity gradient with orbital phase. The resulting magnitude of the velocity amplitude uncertainty is typically $< 5$ \kms\ for a planet simulated with greater than 99.9 per cent significance but may be as much as \hbox{10\,-\,20 \kms}\ where a fake planetary signature is injected with \hbox{$\sim$ 95.4} per cent significance. This is especially true if the data contain systematics above the photon noise level of the data.

\subsection{Standard model}
In Fig. \ref{fig:planet_signal_kband} (top left) we present the phased deconvolved timeseries of the residual spectra (i.e. spectra with removed stellar spectrum and tellurics and containing a potential planetary signature) based on our standard model (see \S 3). For plotting purposes only, the timeseries has been normalised using the formal variances since some phases (particularly $\phi = 0.372\,-\,0.429$ at the end of the first night of observations ) are more noisy than others. This enables the noise structure at all phases to be more clearly seen. The black and white greyscale values are set at $\pm 1.5\sigma$ in the plotted normalised timeseries. It should be stressed that the true formal variances are utilised when searching for the planetary signal in the un-normalised deconvolved timeseries. Hence spectra with lower S/N receive a lower weighting in our analysis. {Since the planet undergoes eclipse during phases $\phi = 0.498\,-\,0.517$, we do no use spectra taken during this interval in our analysis. These spectra are however plotted for completeness in Fig. \& \ref{fig:planet_signal_kband}.}

In Fig. \ref{fig:planet_signal_kband} (top right) we present the \chisq\ landscape plot of log($\epsilon_0$) vs $K_p$ based on our standard model. Dark features in the plot represent enhancements in \chisq\ whose significance can be measured relative to the plotted confidence levels. The large black feature representing the greatest enhancement in \chisq\ appears with low confidence (in the 68.3\,-\,95.4 per cent confidence region) at log($\epsilon_0$) = -3.41 and \hbox{$K_p = 85$ kms$^{-1}$}. However, as can be seen in the phased timeseries, a number of low-level residual features are present. These appear as dark absorption areas, covering localised regions of velocity and phase. We believe that these features are responsible for the \hbox{$K_p = 85$ kms$^{-1}$} signature and result from imperfect removal of telluric and stellar lines during our analysis, giving rise to false signals. As has been demonstrated previously \citep{barnes08}, a clear detection of the planetary signature would be expected to result in a more localised \chisq\ enhancement and greater significance than the \hbox{$K_p = 85$ kms$^{-1}$} \chisq\ enhancement. Nevertheless, $K_p$ is known for \hdb\ (since the system is eclipsing and the orbital inclination is known) and is indicated by the dashed lines in Fig. \ref{fig:planet_signal_kband}. A candidate planetary signature should thus appear at, or close to (see above), this velocity amplitude.

\begin{figure}
\begin{center}
   \includegraphics[width=85mm,angle=0]{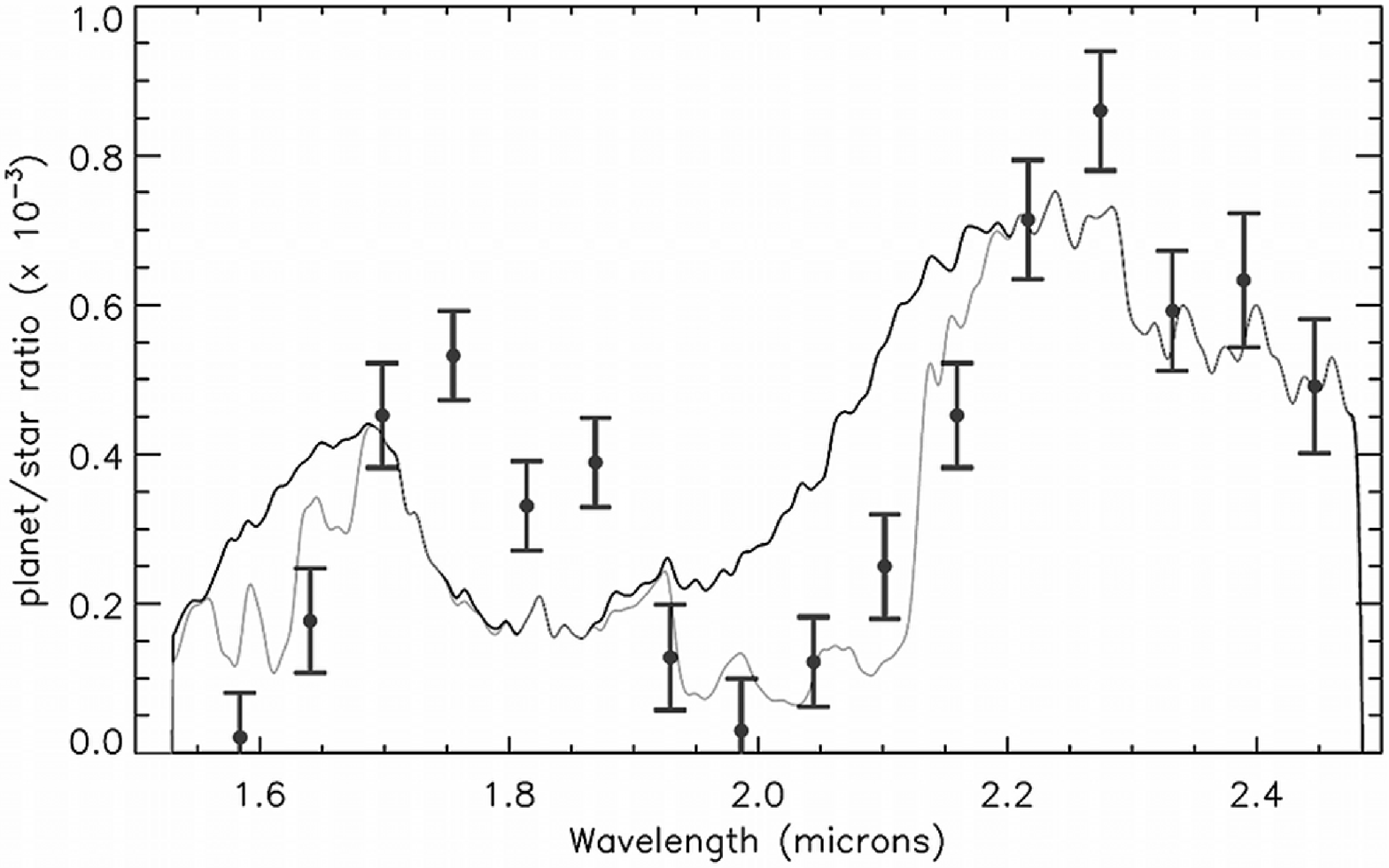} 
   \caption{Comparison of the standard model (black) and augmented CO$_2$ abundance model (grey) with the observed NICOMS/HST dayside spectrum of S09. The enhanced CO$_2$ model shows a considerably improved fit in the 1.9\,-\,2.2 \micron\ region relative to the standard model.}
\end{center}
\protect\label{fig:swain_vs_model}
\end{figure}




For \hdb, the standard model predicts a maximum contrast ratio of \hbox{log$_{10}(\epsilon_0)$ = -3.163} or \hbox{$F_p/F_*$ = 1/1460} over the wavelength span of our observations. We are however unable to detect the planetary signature, at the mean deconvolved wavelength of 2.19 \micron, with 68.3, 95.4, 99 and 99.9 percent confidence levels of \hbox{log$_{10}(\epsilon_0)$ = -4.065,} -3.491, -3.366 \& -3.193 or \hbox{$F_p/F_*$ = 1/11600,} 1/3100, 1/2320 \& 1/1560 respectively. In light of much better observing conditions, this is a significantly more sensitive result than our 2006 observations \citep{barnes07b} permitted. Considerable care must be exercised if quoting sensitivities at contrasts ratios deeper than the 95.4 per cent level (this is investigated further in section \S 4.3 below) owing to candidate signatures which arise from systematics at these levels. HD 189733b is therefore not detected at $K_p =$\ 152.6 \kms\ at a contrast which is 2.1 and 1.1 times deeper (95.4 per and 99.9 per cent confidence respectively) than the standard model predicts.

\begin{figure*}
 \begin{center}
   \begin{tabular}{cc}

      \vspace{10mm} \\
      \includegraphics[width=70mm,bbllx=113,bblly=114,bburx=397,bbury=397,angle=0]{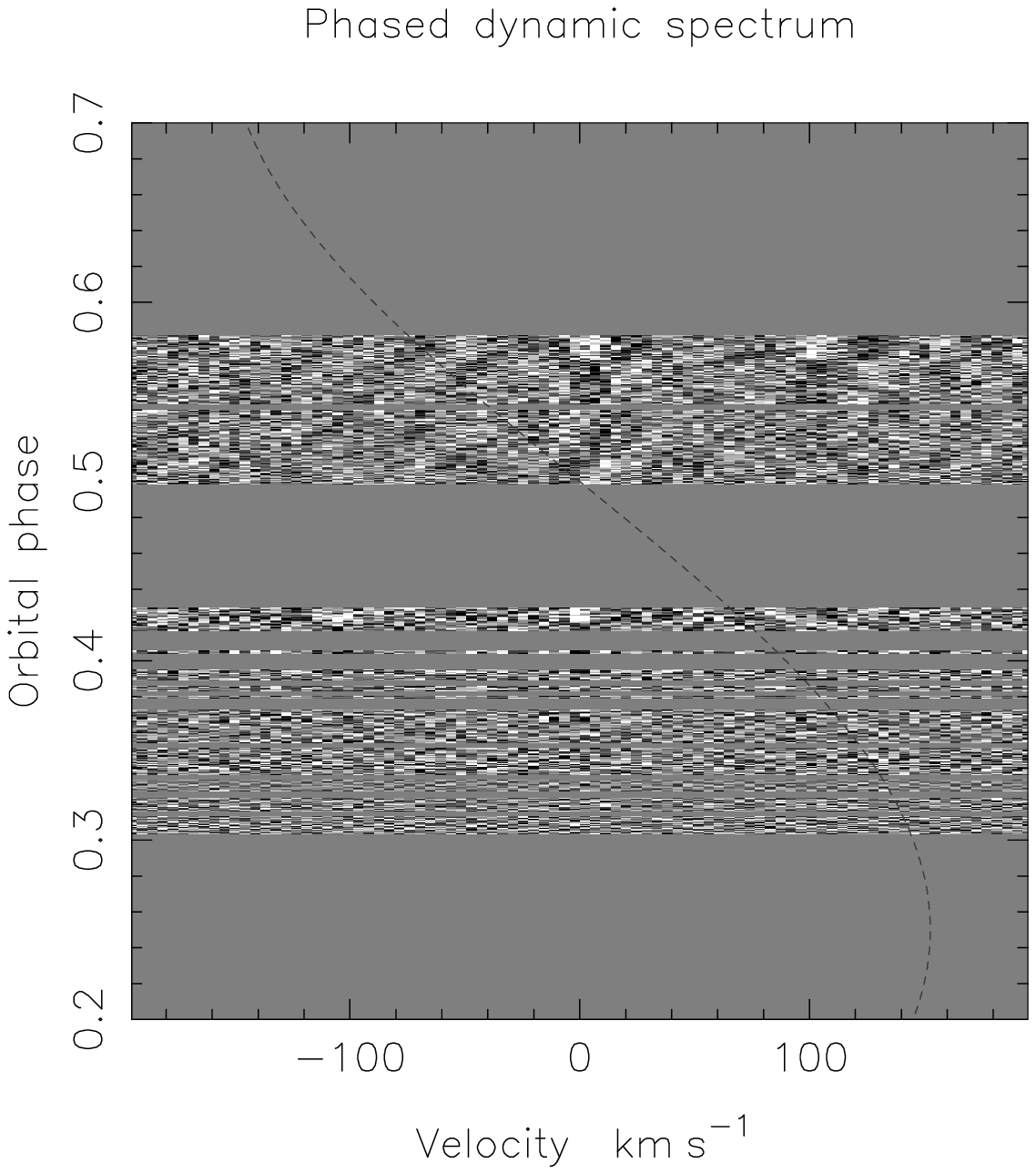} \hspace {5mm} &
      \hspace {5mm}
      \includegraphics[width=70mm,bbllx=113,bblly=114,bburx=397,bbury=397,angle=0]{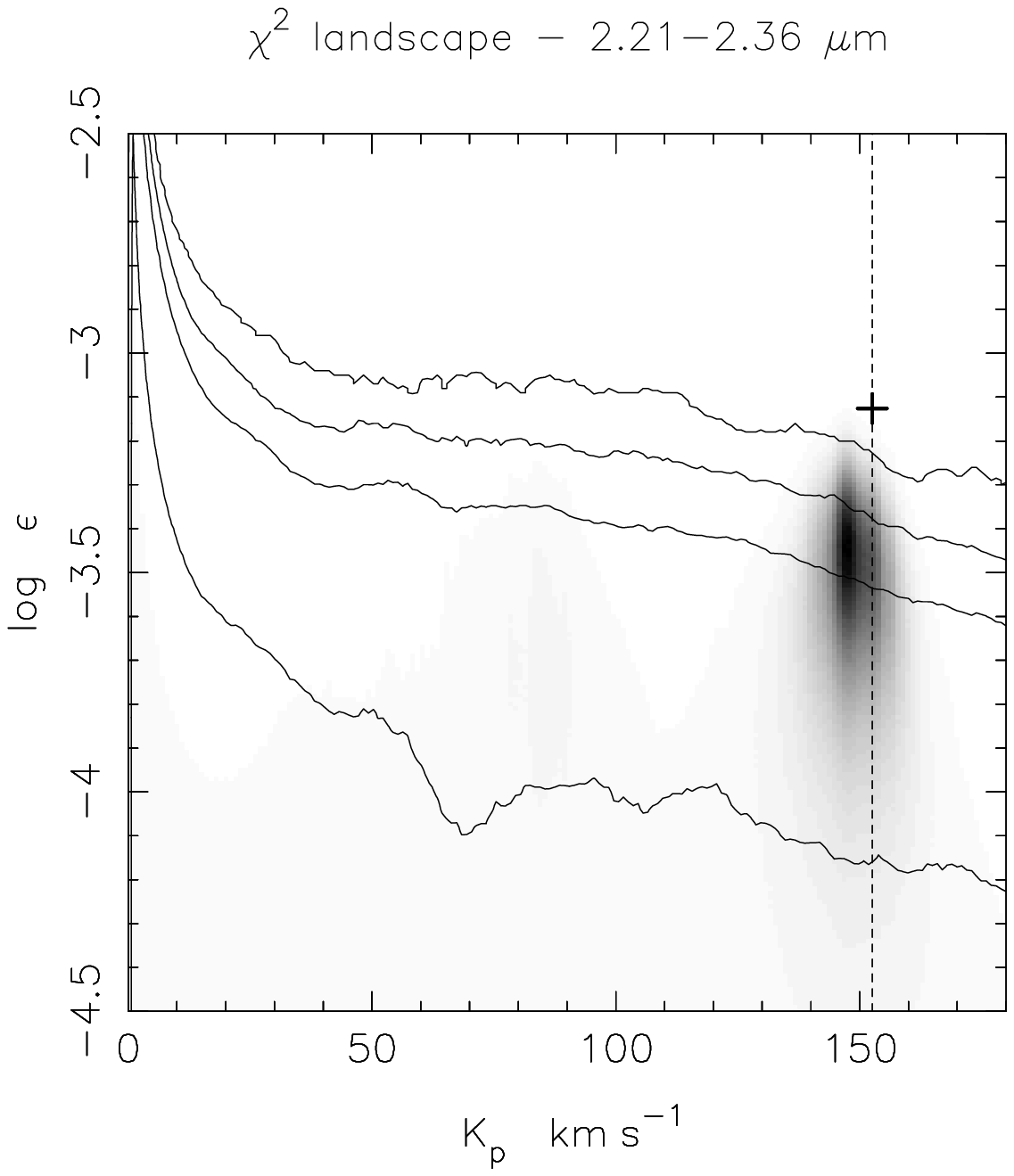} \\
      \vspace{10mm} \\

   \end{tabular}
 \end{center}
\caption{As for Fig. 2, but using only the wavelength range 2.21\,-\,2.36 \micron. The corresponding NICMOS/HST S09 contrast ratio for this region is marked by a + symbol with $F_p/F_* \sim 1/1340$. A candidate signature is detected with 97.2 per cent confidence close to the expected $K_p$ velocity, although the contrast ratio of \hbox{log$_{10}(\epsilon_0)$ = -3.449 ($F_p/F_* = 1/2810$)} is lower than expected. See main text for details.}
\protect\label{fig:split}
\end{figure*}

\subsection{Enhanced CO$_2$ model}
Recent Spitzer/IRAC \citep{grillmair08dayside} and HST/NICMOS (S09) observations have enabled the dayside spectrum of \hdb\ to be measured using low resolution spectroscopy. In order to reliably fit the spectrum, a greater than expected abundance of CO$_2$ is required with S09 reporting CO$_2$ mixing ratios of $c \sim 0.1\,-\,1 \times 10^{-6}$. We have generated a model with augmented CO$_2$ abundance which enables us to match the HST/NICMOS observations (Fig. 3). From here on, in \S 4, we only consider this model. We emphasise that our current model does not contain some of the hot CO$_2$ bands which have been identified by Fourier Transform Spectroscopy carried out at the Jet Propulsion Laboratory and included in the latest edition of HITRAN (see \citealt{rothman09hitran} and references therein). We therefore caution that in order to achieve the required level of absorption from CO$_2$ in the \hbox{1.9\,-\,2.2 \micron,} the relative strengths of individual opacities in the output model are likely overestimated. Nevertheless, based on {the S09} results, inclusion of such opacities would appear to give a more accurate representation of the expected opacities found in the spectrum of \hdb. Fig. 2 (bottom left and bottom right) presents the deconvolved timeseries and log($\epsilon_0$) vs $K_p$ \chisq\ plot after deconvolution using our augmented CO$_2$ model spectrum. At \hbox{$K_p = 152.6$ kms$^{-1}$}, the expected planetary signature is not detected with 68.3, 95.4, 99 and 99.9 per cent confidence levels of \hbox{log$_{10}(\epsilon_0)$ = -4.074, -3.529 \& -3.400 \& -3.283} or $F_p/F_*$ = 1/11600, 1/3380, 1/2510 \& 1/1920 respectively. The sensitivities are slightly greater than for the standard model, although we note that the enhanced CO$_2$ sensitivities are quoted for a centroidal wavelength of 2.15 \micron\ rather than 2.19 \micron. This shift in mean wavelength results from greater {\em normalised} depths of the enhanced CO$_2$ at the shorter wavelengths of our observations. In addition, although the mean planetary flux level is lower in the regions with enhanced CO$_2$, the recorded count rate in the observed star\,+\,planet spectra is higher, leading to higher contrast confidence limits with this model. The mean 2.0\,-\,2.4 \micron\ planet/star flux ratio reported by S09 is \hbox{log$_{10}(\epsilon_0)$ = -3.286} or $F_p/F_*$ = 1/1930 {(i.e. almost identical to our 99.9 per cent confidence level of 1/1920)}, indicating that we are sensitive down to the 99.9 per cent level.




\subsection{Wavelength splitting the data - a candidate signature?}
While there is no clear candidate signature at the expected velocity amplitude of the planet in the enhanced CO$_2$ deconvolved timeseries, we do detect a signal with relatively low confidence (95.4 percent) at \hbox{$K_p = 167.9$ \kms} and \hbox{log$_{10}(\epsilon_0)$} = -3.63 (\hbox{$F_p/F_*$ = 1/4270}). Although it is likely that this signature is again the result of alignment of systematic absorption features in the timeseries at this velocity amplitude, we have investigated splitting the timeseries data into two spectral regions. Deconvolution was carried out on the first three orders (region 1: \hbox{2.03\,-\,2.18 \micron}) and on the second three orders (region 2: \hbox{2.21\,-\,2.36 \micron}) independently before carrying out a search for the planetary signature. With the data split in this manner, region 1 contains H$_2$O and CO$_2$ opacities while region 2 contains H$_2$O and CO (bandhead at $\sim 2.29$ \micron) opacities. While region 1 did not reveal any candidate signature close to \hbox{$K_p = 152.6$ kms$^{-1}$}, region 2 has a more well defined candidate signature at \hbox{$K_p = 147.8$ kms$^{-1}$} with \hbox{log$_{10}(\epsilon_0)$ = -3.449 (1/2810)} and 97.2 per cent confidence. The S09 contrast ratio for \hdb\ over the wavelength span of region 2 is ($F_p/F_* \sim 1/1340$). To assess the true nature of the region 2 signature, we refer the reader to \hbox{Fig. 4 (left)} which again indicates that there are a number of systematic features. Close examination reveals that the expected velocity position of the planet as a function of phase (indicated by the dashed line) appears to pass through, or near, a number of contiguous absorption regions. To investigate the contribution of these regions to the candidate signature, we have carried out three tests.

\medskip
\noindent
1) Analysis of the data observed on 15th June - $\phi = 0.303 - 0.429$ alone. The result is a candidate signature with \hbox{$K_p = 153.9$ kms$^{-1}$} and with \hbox{log$_{10}(\epsilon_0)$ = -3.008 (1/1020)} and 98.8 per cent confidence. \\
2) Analysis of the data observed on 22nd June - $\phi = 0.517 - 0.581$ alone. A candidate signature with \hbox{$K_p = 141.8$ kms$^{-1}$} and with \hbox{log$_{10}(\epsilon_0)$ = -3.528 (1/3372)} and 94.8 per cent confidence is found. \\
3) Contiguous dark regions omitted from the analysis by eye (the regions are not of sufficient amplitude to enable reliable sigma-clipping). No candidate signature within \hbox{26 kms$^{-1}$} of the known \hbox{$K_p = 152.6$ kms$^{-1}$} is apparent. \\ 

The varying velocity and contrast ratio of the candidate signals from tests 1 and 2 suggest that if any planetary signature contributes to the \chisq\ enhancements, it is biased by some other factor. Test 3, in which the contiguous regions are omitted from the analysis, has the effect of removing the candidate signature seen in Fig. 4 (right) completely. The confidence levels, at \hbox{$K_p = 152.6$ \kms}, with the omitted contiguous residual absorption regions are \hbox{log$_{10}(\epsilon_0)$ = -3.964, -3.355 \& -3.214 \& -3.098} or $F_p/F_* = 1/9200, 1/2260, 1/1640$ \& $1/1250$ respectively. In other words, the phases which do not show contiguous blocks of absorption residuals in the timeseries (66 per cent of the recorded spectra) along the radial velocity path of the planet do not possess the ability to recover a planetary signal.

We stress that the argument asserting that the contiguous residual absorption regions are wholly due to systematics and solely responsible for producing candidate signatures is however not strictly true. Any residual absorption features in the timeseries have the ability to modify the contrast ratio and velocity amplitude of a true planetary signal. Since the residual absorption features may be expected to vary in strength it is not unlikely that they would result in a planetary signature modified by differing degrees in tests 1 and 2. In our third test, removing 33 per cent of the data along the expected radial velocity curve of the planet leaves only regions which are consistent with the mean level, or regions of contiguous ``emission'' relative to the mean level. One might expect that this procedure would severely impair our ability to detect a planetary absorption signature. In the hypothesis that dark regions {\em are} artifacts of the data processing (i.e. imperfect telluric/stellar line removal), after their removal \hbox{(test 3)} we can rule out our ability to detect the planetary signal with 95.4 per cent confidence at a level of $F_p/F_* = 1/2260$. In light of this and our 99.9 per cent upper limit (\S 4.2) on the contrast ratio, we believe that since S09 detect the planet with $F_p/F_* = 1/1930$, further investigations of model dependency on our analysis are required.

\section{Model dependency effects}

We are confident that the planetary signature is not severely attenuated (there is inevitably some attenuation as described in \S 4) during our analysis procedure since fake planetary signals which are injected before analysis are recovered. A cause of our inability to detect a planetary signature is likely to stem from a mismatch between the model planetary spectrum and observed planetary spectrum. The most likely direct causes of line strength mismatch and model line wavelength opacities were first highlighted in \cite{barnes07a}. Line strength mismatches may arise from incorrect treatment of the model atmosphere, including uncertainty in the exact form of the temperature-pressure (T-P) profile. In addition, the precision of the calculated opacities is limited by the accuracy of the Einstein A coefficients. This latter effect may be true for important molecular species such as H$_2$O \citep{barber06water} for instance. We investigate relative line depth, temperature and wavelength uncertainties below.

\subsection{Relative line depths}

Although the model planetary spectrum may show little variation as a result of T-P profile changes when observed at low resolution, the relative line depths may change significantly. In addition, the gradient of the T-P profile determines the absolute strength of the absorption lines. To investigate these effects, we have generated a series of ad hoc T-P profiles and resulting emergent spectra. Fig. 5 shows the models plotted for a short region of wavelength space. Steeper T-P gradients lead to the formation of deeper lines whereas the relative line strengths vary from model to model. These effects are important since mismatch of the model and observed spectra line depths will lead to a non-optimally deconvolved line profile, and therefore decrease in sensitivity. If all the lines are less deep, they will simply be harder to detect above the noise level.

To investigate the degree to which our ad hoc models affect the sensitivity of our procedure we used our standard model (black in Fig. 5) to inject a fake planet into the \hd\ timeseries. The fake planet was then recovered and calibrated to match the contrast ratio at which it was injected. By deconvolving with line lists derived from each of the different models shown in Fig. 5, we find that we are able to recover the planetary signature in all cases. The planetary signature is however recovered with an incorrect contrast ratio and modified relative confidence. Fig. 6 shows the relative contrast ratio for a fake \hdb\ planet injected into the data with 99.9 per cent significance. Model 0 represents the standard model calibration to which the simulations are normalised. The contrast ratio is incorrectly recovered, with model 1 showing a 2.3 per cent overestimation of the contrast level and model 4 indicating a 66 per cent underestimation of the contrast ratio. In all cases however, the ad hoc models appear to recover the planet with increased significance. The effect is nevertheless relatively small, with models 2 and 3 showing the greatest increase in confidence. Model 2 indicates an increase in significance of 12.5 per cent relative to the 99 per cent confidence level. One might naturally expect a decrease in confidence to arise from mismatch of the line strengths during deconvolution rather than the counter-intuitive increase. We believe that the increase is most likely due to models 1\,-\,4 yielding strong lines which become even stronger and weak lines which become weaker relative to the standard model. If one of these models were a closer match to the empirical \hdb\ spectrum, we note that the relative significance of the standard model would decrease (with a maximum reduction in sensitivity of 14.8 per cent relative to the 99 per cent confidence level). In conclusion, the above ad hoc models alone are not able to explain the lack of true planetary signal (our observations are after all still sensitive enough to detect \hdb) through mismatch of line strengths. Relative to the confidence levels in Fig. 6, a true planetary signal could not change its confidence by more than half of the separation of the 99 and 99.9 per cent confidence levels.

\begin{figure}
\begin{center}
   \includegraphics[width=84mm,angle=0]{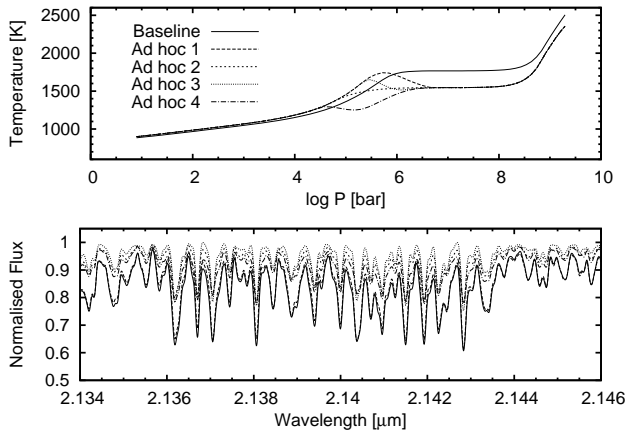} \\
   \caption{Ad hoc T-P profiles and resulting spectra for 2.135\,-\,2.145\micron\ region. The black profile and spectrum represents the standard, standard model. While a steeper temperature profile at pressures where lines predominantly form (e.g. red/blue models) result in spectra with deeper absorption lines, note that the relative line strengths also differ from model to model.}
\end{center}
\protect\label{fig:tp_profiles_spectra}
\end{figure}

\subsection{Opacity wavelength uncertainties}

Moderate wavelength uncertainties lead to an effective degradation of deconvolved resolution while model temperature uncertainties may lead to line strength mismatches with the observed spectra during deconvolution. We were able to use an unpublished improved version of the BT2 \citep{barber06water} water line list (with reduced wavelength uncertainties) to investigate these effects. Most of the strong lines which contribute to the deconvolution are transitions between states whose energies are experimentally known to very high accuracies. Consequently, by selecting only those lines which are greater in strength than 1/10,000 of the strongest line, we eliminate a large number of lines which are expected to have larger uncertainties in calculated positions and which in any case have negligible contribution to the deconvolved profile. Where the energies of the upper and lower states of a transition are both known experimentally, these are used for modified BT2 line frequencies, rather than the ab initio calculated values. At 1250 K, 75 per cent of the water lines in our trimmed list are transitions between experimentally-known levels, and only 25 per cent employ BT2 ab initio frequencies/wavelengths (in all cases, however, BT2 Einstein A values are used in computing line {\em strengths} as these are generally more accurate than experimentally-determined values). 

\cite{barber06water} state that a comparison of the BT2 ab initio frequencies with experimentally-known transitions shows that the positions of $\sim 40$ per cent of the lines tested are accurate to within 0.1 cm$^{-1}$ and 91 per cent are within 0.3 cm$^{-1}$. At 2.2 \micron, this corresponds to resolutions of R = 45,500 and R = 15,150 respectively. Clearly, since water is the dominating opacity, these uncertainties will play a role in degrading the resolution of a deconvolved profile for data sets with resolutions of R\,$\geq$\,15,000. These uncertainties are therefore applied to the 25 per cent of ab initio lines in our 1250 K list by using a Gaussian random uncertainty. This should represent a worst case scenario because we have removed those lines which are weaker than 1/10,000 of the strongest line and which are expected to exhibit the largest frequency/wavelength errors. We then carry out simulations by injecting a planetary signature into the \hdb\ spectra using our 1250 K spectrum and deconvolving firstly with the matching line list (case A), and the with an adjusted line list which models the wavelength uncertainties (case B). The mismatch (i.e. case B relative to case A) leads to a planet which is detected with a 6.5 per cent underestimation of contrast ratio and a 14.5 per cent decrease in significance relative to the 99.9 per cent confidence level.

\subsection{Temperature uncertainty}

The effect of using a model line list for deconvolution which varies from the observed spectrum in temperature alone is shown in Fig. 7. Here, a \hbox{1250 K} planetary spectrum signature is recovered with 99.9 per cent confidence using a \hbox{1250 K} modified BT2 line list. However both the recovered contrast ratio and significance change when deconvolved with line list temperatures which differ by $\pm$250 K and $\pm$500 K from 1250 K. The effect is again relatively small for an underestimation of temperature (the increase in significance is likely due to over-weighting of strong lines and under-weighting of weak lines as described in \S 5.1) while slightly more significant for overestimation of temperature. In all instances a planetary signature is however recovered. For HD 189733b, the above effects alone are not sufficient to explain the lack of planetary signature (using our standard model) which is predicted at the log$_{10}(\epsilon_0)$ = -3.163 level. Combined wavelength and temperature mismatches should lead to a 99.9 per cent planetary signature appearing with 95.4 per cent confidence at worst. A $\pm 250$K mismatch in model spectrum temperature results in a 20.5 per cent relative uncertainty in the confidence of a recovered signal.

\begin{figure}
\begin{center}
   \includegraphics[width=60mm,angle=270]{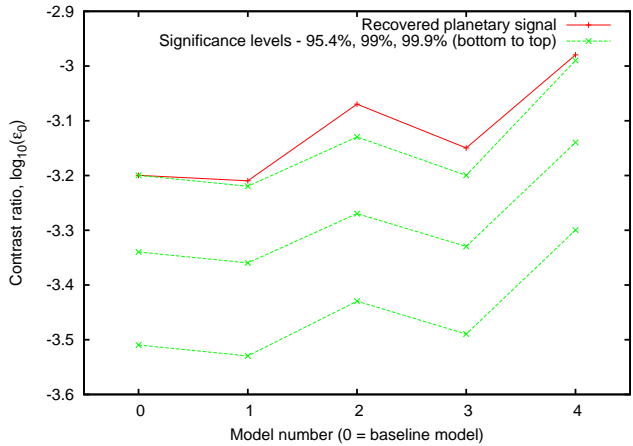}
   \caption{Relative contrast ratio vs ad hoc deconvolution model. The standard model is represented by model 0. A fake planet was injected into the \hdb\ timeseries with 99.9 per cent confidence and recovered using our procedure. For each of models 1\,-\,4, the deconvolution step was carried out using the corresponding model. The solid line represents the recovered planetary signature relative to the standard model 0. The dashed lines are the 95.4, 99 and 99.9 per cent confidence levels (bottom to top) respectively. Note how the contrast ratio is incorrectly recovered with models 1\,-\,4, while the relative significance increases relative to the standard model.}
\end{center}
\protect\label{fig:strength_comparison}
\end{figure}

\begin{figure}
\begin{center}
      \includegraphics[width=60mm,angle=270]{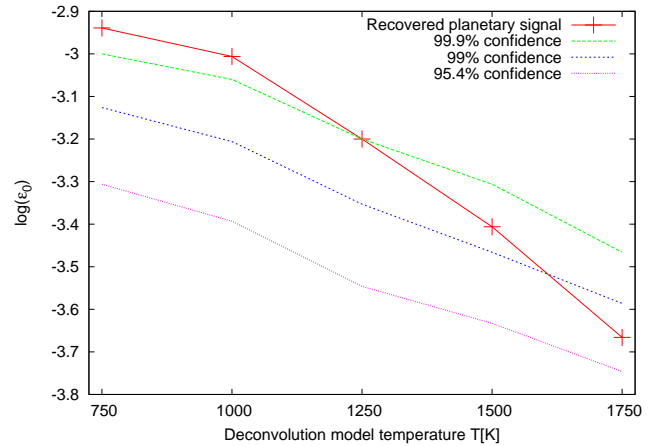} \\
  \caption{Significance of recovered planet as a function of temperature. A planet with 99.9\% significance is simulated with a temperature of 1250K and recovered with 99.9\% significance. By deconvolving the spectra with a range of temperatures, the recovered contrast ratio and significance of the planet are seen to vary.}
\end{center}
\protect\label{fig:tempmismatch}
\end{figure}

\subsection{Other sources of uncertainty}

Additional model and observational uncertainties may contribute to an incorrect estimate of the planet/star contrast ratio or its relative significance. A possible source of error may arise from the planet ephemeris although this has been determined to high precision. Following our previous study of the \hd\ system \citep{barnes07b}, we adopt the ephemeris of \cite{winn07hd189733b}, determined from Stromgen b and y passband observations ($T_t = 2453988.80336(23) + 2.2185733(19)$). A more recent estimate of the ephemeris by \cite{agol08} using Spitzer 8 \micron\ observations of planetary transit yields T$_t = 2454279.436741(23) + 2.21857503(37)$. Since limb darkening and starspot effects are reduced at longer wavelengths, this has been claimed as the most precise measurement of the ephemeris to date. The predicted mid-transit time for our observations differs by 73 secs when comparing the two ephemerides, a phase difference of 0.00038. The level of precision of ephemeris observations from \hdb\ is therefore now sufficient that new refinements have no measurable effect on the contrast or velocity amplitude of a planetary signal.

A more important consideration arises from global re-circulation patterns in the atmosphere of \hdb. For a tidally locked planet with a static atmosphere, one might expect the maximum planet/star contrast ratio to occur at orbital phase $\phi = 0.5$ due to the highest effective irradiation of the planet at the sub stellar point. \cite{knutson07hd189733b} found a difference in day and night side brightness temperatures of 238 K from 8 \micron\ photometric Spitzer light curve variations. The $1212 \pm 11$ K dayside temperature was found to be displaced from the substellar point by $16 \pm 6$\degs. This finding is in accordance with 3D circulation models \cite{showman02,cooper05,fortney06,showman08}. More recently, \cite{knutson09} have re-analysed their 8 \micron\ lightcurves and published 24 \micron\ lightcurves of \hdb. Maximum brightness is found to occur at phase $0.396 \pm 0.022$ corresponding to a shift eastward of 20\degs\,-\,30\degs of the hottest region relative to the substellar point. We have carried out a simulation to estimate the effect of such a shift which is not accounted for in the preceding analysis. We created an artificial planetary signal which peaked 30\degs\ before secondary eclipse and recovered with a phase function which peaked at the same shift {\em and also at} secondary eclipse. The recovered planetary signature which did not account for the 30\degs\ shift was found to overestimate the contrast ratio by $\sim 10$ per cent. The relative change in significance increases by 4.7 per cent since the contrast ratio must be increased to optimise the \chisq\ fit to the mis-aligned phase function. We note that this effect will be dependent on observational phase coverage and S/N ratio from night to night (i.e. shifting the phase function peak to a region of fewer observations or lower S/N ratio will reduce sensitivity).

We have assumed an effective \vsini\ for \hdb\ of 2.53 \kms\ which corresponds to a tidally locked planet. However there may be additional broadening as a result of the re-distribution of heat. \cite{showman08} find that up to 3-4 \kms\ wind speeds are responsible for the advection of heat away from the substellar point. This shift is somewhat less than our resolution element of 11.99 \kms. Although the wind speeds are effectively translational (an east-west flow) at the 100-1000 mbar levels from which the 2.2 \micron\ spectrum is expected to predominantly arise (see Fig. 4 of \cite{showman08}), we have simulated an additional 4 \kms\ broadening of the spectral lines. Combined in quadrature with the rotational broadening, we simulate a planetary atmosphere which possesses lines broadened by 4.73 \kms\ rather than 2.53 \kms\ from rotational broadening alone as in the preceding sections. As expected this effect is also minor at a resolution of 25,000 with a 9.6 percent drop in sensitivity.

\subsection{A Semi-empirical approach - an L dwarf spectrum}

In addition to model uncertainties, we have carried out a semi-empirical examination of our ability to recover the spectrum of a brown dwarf which closely matches the planetary temperature of 1250 K. K band observations of an L3.5\,-\,L4.5 spectrum \citep{kirkpatrick00,knapp04} were secured by with NIRSPEC at a spectral resolution of 22,000, covering redder wavelengths in each order than the \hdb\ observations. We were thus unable to use the L spectrum as a template which could be injected into our timeseries to mimic the signature of a fake ``planet''. Instead, using the Taylor expansion scaling technique described in \S 2.3, we scaled the standard \hdb\ model spectrum to give the closest possible match to our observed L spectrum. Being an L dwarf, our spectrum exhibits significant rotation, with \vsini\ = 32 kms$^{-1}$ \citep{osorio06,reiners08activity}. The same broadening was applied to our standard model prior to scaling it to the L spectrum. Deconvolution was then carried out using: (a) the standard model line list on the scaled standard model spectrum and, (b) the standard model line list on the L spectrum. Closely matched deconvolved profiles are recovered in both instances but with a smaller equivalent width for case b. Since \vsini\ is matched, the resulting profiles essentially differ in their depths only, with the case a profile being 55 per cent deeper than case b profile. It is difficult to assess wavelength mismatch effects given the broad nature of the profile; however, we can attribute the 55 per cent decrease in profile strength to line strength mismatches. Although there may be differences between a L spectrum and a planetary spectrum, this semi-empirical approach may be taken to represent an upper limit to our line depth sensitivity. The line depth, wavelength and temperature uncertainties in \S5.1, 5.2 \& 5.3 yield a 28 per cent reduction in sensitivity when combined in quadrature. The semi-empirical analysis result may be equated with this combination of effects, and is almost twice the modelling estimate.

\section{Summary \& Discussion}

We have carried out a high resolution search for the signature of the close orbiting extrasolar giant planet, \hdb. Our signal enhancement technique enables us to achieve the sensitivities required to detect the dayside spectrum of the planet that has already been observed at a mean contrast ratio of $F_p/F_* = 1/1930$ by S09 in the K band region of our observations. Inclusion of augmented CO$_2$ abundance is however not sufficient to detect the planet with a 99.9 percent confidence level of $F_p/F_* = 1/1920$ {(i.e. almost identical to the S09 result)}. A tentative candidate planetary signature {\em is} found at 15 kms$^{-1}$ greater than the expected velocity amplitude of the planet at $F_p/F_* = 1/4270$. In light of the model uncertainties that have been investigated, finding a planetary signature with modified contrast ratio and velocity amplitude {is reasonable}. This prompted us to perform simulations in which planetary signals were injected at contrast ratio levels equivalent to those induced by contiguous absorption residuals. While these planetary signatures could be recovered, we found that the velocity amplitude may be uncertain by $\pm 20$ kms$^{-1}$, further reflecting the difficulty of reliably extracting a real signal at the 95.4 per cent level. We note however that a planetary signature with 99.9 per cent confidence should easily be detected, as demonstrated in \cite{barnes08}. 

Splitting the data into two wavelength regions revealed that the 2.21\,-\,2.36 \micron\ region (containing mainly H$_2$O and CO opacities) yielded a candidate planetary signature with higher confidence. Analysing these subsetted data on a night by night basis however revealed that the signature was not stable in velocity amplitude or contrast ratio suggesting that it could result from a chance alignment of a number of systematic contiguous absorption residual features at the phase dependent velocity position of the planet. By removing these features we found that the signature, close to the known \hbox{$K_p = 152.6$ kms$^{-1}$}, disappeared. The remaining 66 per cent of the data did not possess the ability to recover a planetary signature at the level determined by the results of \cite{swain09dayside} with between 95.4 and 99 per cent confidence. Since the remaining data contained contiguous regions with levels above the mean, this may not be surprising as we only search for absorption signatures. It is important to emphasise that the tests we have carried out do not rule out the possibility that a true planetary signal is contained within the spectra. The detected candidate features may be partially influenced by a true planetary signature, but at the 95.4 per cent levels, no confident claim for a detection can be made.

The effects of model opacity strength uncertainties, wavelength uncertainties, temperature mismatch, phase function mismatch and velocity field/broadening uncertainties contribute sensitivity uncertainties of 12.5, 14.5, 20.5, 4.7 and 9.6 per cent respectively. Combining these effects in quadrature yields a total uncertainty in the significance of the result of 30 per cent. Further, if we take the semi-empirical 55 per cent uncertainty as an upper limit to our line depth, wavelength and temperature mismatches, the corresponding uncertainty is 56 per cent. Assuming that the L spectrum can provide a close match to that of \hdb, the semi-empirical result already shows that the model uncertainties may be significantly underestimated. Hence the 99.9 percent confidence with which we reject a signal at the know $K_p$ could in fact be modified to a level with reduced significance, taking a candidate signal to contrast ratios that are plagued by systematic features.

While the current generation of models can adequately fit broadband photometric and low resolution spectroscopic observations, it is clear that moving to higher resolution requires further model refinement. With the uncertainties investigated above, we can not rule out the presence of the planet using our technique, especially if further model uncertainties remain unaccounted for. Only further observations which would bring about an increase in sensitivity, or more precise model atmospheres could increase our chances of detecting \hdb.

\section{Acknowledgments}
JRB was supported by a STFC funded research grant during the course of this work. TB acknowledges support from NASA's Origins of Solar System program and the NASA Advanced Supercomputing facility, and LP from NSF grant 04-44017. The authors wish to recognise and acknowledge the very significant cultural role and reverence that the summit of Mauna Kea has always had within the indigenous Hawaiian community.  We are most fortunate to have the opportunity to conduct observations from this mountain. We would like to thank the referee for providing constructive suggestions for improving the manuscript.


\end{document}